\documentclass[aps,prd,reprint,superscriptaddress]{revtex4-1}

\usepackage{hyperref}
\usepackage{graphicx}
\usepackage{enumerate}
\hypersetup{
    pdfnewwindow=true,      
    colorlinks=true,       
    linkcolor=blue,          
    citecolor=blue,        
    filecolor=blue,      
    urlcolor=blue           
}

\usepackage{subfigure}
\usepackage{amsmath}
\usepackage{amssymb}
\usepackage{amsfonts}
\usepackage{float}
\usepackage{color}

\providecommand{\apj}[0]{ApJ}
\providecommand{\apjl}[0]{ApJ Lett.}

\providecommand{\aap}[0]{A\&A}

\providecommand{\araa}[0]{Ann.\ Rev. Astron. Astroph. }

\providecommand{\jcap}[0]{J. Cosmol. Astropart. P.}
\providecommand{\mnras}[0]{MNRAS}

\providecommand{\nat}[0]{Nature}

\providecommand{\prd}{PRD}

\begin{document}

\title{Spectral Decline of PeV Neutrinos from Starburst Galaxies}

\author{I. Bartos}
\email{ibartos@phys.columbia.edu}
\affiliation{Department of Physics, Columbia University, New York, NY 10027, USA}
\author{S. M\'arka}
\affiliation{Department of Physics, Columbia University, New York, NY 10027, USA}

\begin{abstract}
Starburst galaxies represent one of the most plausible origins of the cosmic high-energy neutrino flux recently discovered by IceCube. At $\sim$\,PeV energies, the neutrino flux from starburst galaxies is expected to exhibit a characteristic spectral break due to cosmic-rays escaping the galaxy. We examine the 'smearing' of this spectral break by a population of starburst galaxies with varying properties. We incorporate galaxy distribution w.r.t. star-formation rate and redshift. Our results (i) show characteristic spectral softening in IceCube's energy band; (ii) resolve the conflicting observations of soft neutrino spectrum and diffuse gamma-ray flux observed by Fermi-LAT; (iii) constrain the properties of the magnetic fields in starburst galaxies.
\end{abstract}


\maketitle


\section{Introduction}

IceCube has recently discovered a quasi-diffuse flux of astrophysical neutrinos \cite{2013Sci...342E...1I,2015arXiv150703991I,2015arXiv150704005I}. The origin of these neutrinos is currently unknown. Unraveling the source will help understand the origin of cosmic rays, the mechanisms behind particle acceleration, and the broader environment of the acceleration site.

The astrophysical neutrino flux has been observed in the $20$\,TeV - 3\,PeV range to be $\sim 10^{-8}$\,GeV\,cm$^{-2}$s$^{-1}$sr$^{-1}$ per neutrino flavor \cite{2014PhRvL.113j1101A,2015arXiv150703991I}. The spectrum of the flux is, however, currently uncertain. Observations seem to be inconsistent with $dN_{\nu}/dE_\nu\propto E^{-\Gamma}$, with $\Gamma=2$ spectrum typically expected from Fermi acceleration \cite{2015arXiv150703991I,2014PhRvL.113j1101A}. Constraints using muon neutrinos from the Northern hemisphere detected by IceCube between 2010-2012 in the energy range 330\,TeV-1.4\,PeV indicate an effective spectral index of $\Gamma_{\rm eff} \approx 2.2$ \cite{2015arXiv150704005I}. Using a combination of IceCube searches with neutrinos observed in the 25\,TeV-2.8\,PeV energy band, Aartsen et al. \cite{2015arXiv150703991I} finds a best fit spectral index of $\Gamma_{\rm eff} = 2.50\pm0.09$. Results are also compatible with a harder spectrum with a spectral break. Such spectral features are plausible for a variety of source types, including starburst galaxies \cite{2015ApJ...806...24S,2015ApJ...805...95C} or merger shocks in clusters of galaxies \cite{2008ApJ...689L.105M,2009ApJ...707..370K}.

Starburst galaxies represent one of the prime candidates for the origin of high-energy neutrinos \cite{2003ApJ...586L..33R,2006JCAP...05..003L,2007ApJ...654..219T,2011ApJ...734..107L,2014JCAP...09..043T,2015ApJ...805...95C}. High star formation rate (SFR) is connected to the occurrence of phenomena such as gamma-ray bursts \cite{2012ApJ...744...95R} and supernova remnants \cite{1999A&A...351..459C} which are plausibly connected to the acceleration of cosmic rays \cite{1995PhRvL..75..386W,2002Natur.416..823E}. Even more importantly, strong magnetic fields and high interstellar gas densities in starburst galaxies make them largely opaque to most cosmic rays \cite{2006JCAP...05..003L,2007ApJ...654..219T,2011ApJ...734..107L}. Cosmic rays produced within starburst galaxies lose their energy through nuclear collisions, producing high-energy neutrinos. Only the highest energy cosmic rays can escape since they are less affected by magnetic fields in the galaxies. 

\begin{figure}[H]
\begin{center}
\resizebox{0.49\textwidth}{!}{\includegraphics{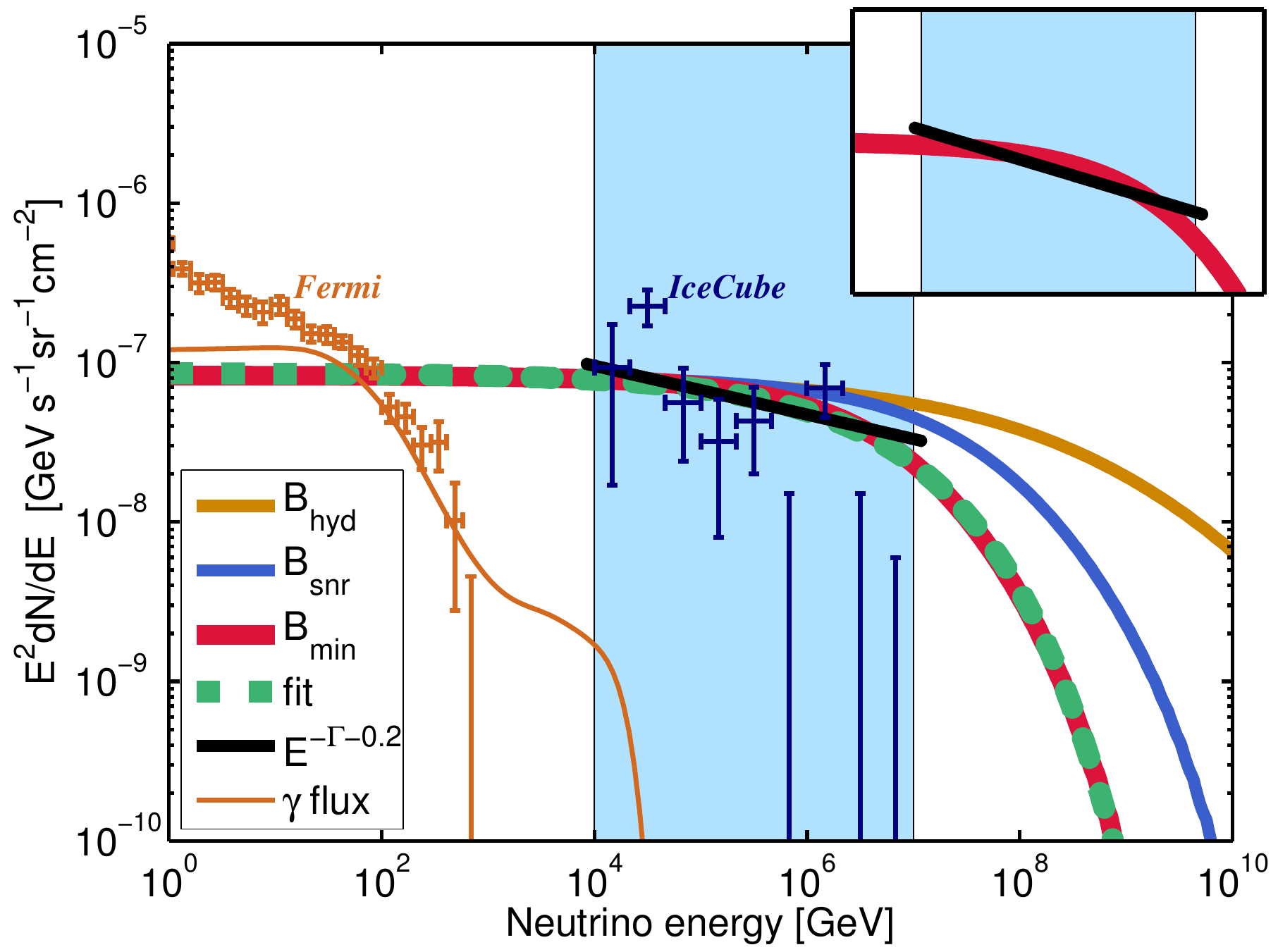}}
\caption{Expected quasi-diffuse neutrino spectrum from starburst galaxies, adopting injection index $\Gamma=2$. Results are shown for three different magnetic field models (see legend), $B_{\rm min}$ being our benchmark model. Also shown is an empirical fit onto the benchmark spectrum (dotted line), and a power law fit onto the 10\,TeV-10\,PeV energy interval (black), which is the energy band of astrophysical neutrinos detected by IceCube. Additionally shown are the observed IceCube and Fermi quasi-diffuse fluxes, as well as the $\gamma$ flux expected given the benchmark neutrino spectrum. The inserted plot separately shows the benchmark spectrum and power-law fit.}
\label{figure:spectrum}
\end{center}
\end{figure}

An additional constraint on the neutrino spectrum comes from the quasi-diffuse GeV gamma-ray flux observed by Fermi-LAT \cite{2010PhRvL.104j1101A}. If cosmic neutrinos are produced in hadronuclear interactions, as expected for the case of starburst galaxies and many other models, there is a corresponding gamma-ray emission that represents an independent probe of the neutrino spectrum. With the latest IceCube observations, the observed Fermi-LAT quasi-diffuse gamma-ray flux constrains the neutrino spectral index to $\Gamma \lesssim 2.1-2.2$, given that the neutrinos are produced in $pp$ interactions \cite{2013PhRvD..88l1301M}. This constraint holds even for a broken power law spectrum. Together with the current best fits on IceCube observations, requires either (i) at least partly photo-hadronic neutrino production or (ii) deviation from single power law spectrum.

In this paper we study the neutrino spectrum expected from starburst galaxies, accounting for their varying properties and cosmic distribution. Our goal is to gain a more detailed picture of the spectral decline due the cosmic ray escape energy, which varies between starburst galaxies. After reviewing the expected neutrino spectrum from a single starburst galaxy, we use SFR to recover corresponding typical values for galaxy properties. We then consider the cosmic distribution of starburst galaxies to derive the cumulative neutrino spectrum. We finally compare the recovered deviations from the power law spectrum expected from the cosmic ray sources within galaxies with neutrino and gamma-ray observations.

\section{Neutrino spectral cutoff in a single starburst galaxy}

We define the neutrino spectrum expected from a single starburst galaxy following \cite{2006JCAP...05..003L}. Below a threshold energy $E_{\rm th}$, the spectrum of cosmic rays in the starburst galaxy is determined by their production spectrum, which we assume to follow a power law with spectral index $p$, and the confinement time $\tau_{\rm conf}$. In the Milky Way,
\begin{equation}
\tau_{\rm conf} \approx 10^7 \left(\frac{E_{\rm p}}{10\,\mbox{GeV}}\right)^{-s} \, \mbox{yr}
\label{eq:conf1}
\end{equation}
for cosmic ray energy $E_{\rm p}$, with $s\approx0.6$ \cite{2006JCAP...05..003L}. 

The threshold $E_{\rm th}$ is determined by the relation between the confinement time and the energy loss time of protons, assuming that the starburst lifetime is sufficiently long \cite{2006JCAP...05..003L}. The energy loss time for hadronuclear interactions is $\tau_{\rm loss} \approx (0.5 n \sigma_{\rm pp} c)^{-1}$, where $n$ is the interstellar nucleon density, and the inelastic nuclear collision cross section $\sigma_{\rm pp}\approx 50$\,mb for the energy range of interest. The factor 0.5 is the collision's inelasticity. Expressing $\tau_{\rm loss}$ with gas surface mass density $\Sigma_{\rm g}\sim m_{\rm p} n h$, $h$ being the galactic disk height, we have
\begin{equation}
\tau_{\rm loss} = \frac{2m_{\rm p} h}{\sigma_{\rm pp} c \Sigma_{\rm g}} = 7\times 10^{4} \left(\frac{\Sigma_{\rm g}}{\mbox{g}\,\mbox{cm}^{-2}}\right)^{-1}\left(\frac{h}{h_G}\right)\,\mbox{yr},
\end{equation}
where $h_G\sim300$\,pc is the Galactic disk height. For the confinement time, we start with Eq. \ref{eq:conf1}, and assume that it depends on the cosmic ray energy only through the Larmor radius ($\propto E/B$). We further assume that the confinement time scales as the square of the galactic disk height since this is a diffusion process. We arrive at
\begin{equation}
\tau_{\rm conf} \approx 10^7 \left(\frac{E_{\rm p}}{10\,\mbox{GeV}}\right)^{-s} \left(\frac{B}{B_{G}}\right)^{s} \left(\frac{h}{h_G}\right)^2 \, \mbox{yr}
\end{equation}
where $B_G\sim6\,\mu$G is the Galactic magnetic field strength.

We can further utilize the connection between $B$ and $\Sigma_{\rm g}$. Primarily, $B$ is estimated using the observed synchrotron emission of cosmic-ray electrons, although there are indications that the magnetic fields in starburst galaxies can be significantly stronger than this estimate \cite{2006ApJ...645..186T,2014ApJ...780..182M}. We follow McBride et al. \cite{2014ApJ...780..182M} and denote this magnetic field estimate based on synchrotron emission as $B_{\rm min}$, since this represents a minimum strength. This will be our benchmark model. Based on McBride et al. \cite{2014ApJ...780..182M}, $B_{\rm min}$ can be expressed as
\begin{equation}
B_{\rm min} = 10^{-4}\left(\frac{\Sigma_{\rm g}}{\mbox{g\,cm}^{-2}}\right)^{0.4}\,\mbox{G}.
\end{equation}
Magnetic field strengths can also be derived from the radio luminosities of supernova remnants. We will denote this estimate with $B_{\rm snr}$. Another possibility is that the magnetic energy density is equal to the pressure of a self gravitating disk in hydrostatic equilibrium, as a function of gas surface density. We will denote the estimate based on this connection with $B_{\rm hyd}$. For these two other possibilities the conversions according to McBride et al. \cite{2014ApJ...780..182M} are $B_{\rm snr} = 6 \times 10^{-4}(\Sigma_{\rm g}/\mbox{g\,cm}^{-2})^{0.55}\,$G and $B_{\rm hyd} = 62 \times 10^{-3}(\Sigma_{\rm g}/\mbox{g\,cm}^{-2})\,$G.

Using these conversions, we can now determine $E_{\rm th}$ by finding the energy for which $\tau_{\rm loss}=\tau_{\rm conf}$. For our benchmark model with $B_{\rm min}$, we find
\begin{equation}
E_{\rm th}(B_{\rm min}) = 6.5\times10^2\left(\frac{\Sigma_{\rm g}}{\mbox{g\,cm}^{-2}}\right)^{2.1}\left(\frac{h}{h_G}\right)^{1.7}\,\mbox{TeV}.
\end{equation}
For the other two models, we find $E_{\rm th}(B_{\rm snr}) = 4\times10^3(\Sigma_{\rm g}/\mbox{g\,cm}^{-2})^{2.2}(h/h_G)^{1.7}\,$TeV and $E_{\rm th}(B_{\rm dyn}) = 6\times10^4(\Sigma_{\rm g}/\mbox{g\,cm}^{-2})^{3.3}(h/h_G)^{1.7}\,$TeV. Note that $E_{\rm th}$ is the threshold for cosmic rays; the corresponding threshold in the neutrino spectrum is $\sim 0.05 E_{\rm th}$ since neutrinos are produced with about $5\%$ of the proton energy in interactions.

\section{Starburst galaxy characteristic parameters from star-formation rate}

We saw above that the spectral features of a starburst galaxy can essentially be expressed with one parameter: $\Sigma_{\rm g}$. The weight of each galaxy in an ensemble spectrum, however, also needs to be taken into account. Further, there is no readily available catalog or model for the cosmic distribution of $\Sigma_{\rm g}$ for starburst galaxies. To overcome this, here we derive a connection between SFR and $\Sigma_{\rm g}$. SFR can be used as a measure of the cosmic-ray emission from a starburst galaxy. It is also more straightforward to observationally determine than $\Sigma_{\rm g}$. The cosmic distribution of starburst galaxies w.r.t. SFR has been characterized \cite{2012ApJ...747L..31S}.

Starting with the SFR of a starburst galaxy (determined, e.g., from its far infrared emission \cite{1998ARA&A..36..189K}), we first convert SFR to SFR surface density $\Sigma_{\rm SFR}$. For galactic half-light (or effective) radius $R$, we define $\Sigma_{\rm SFR}= \mbox{SFR} / \pi R^2$ (e.g., \cite{2010MNRAS.407.2091G}).

To determine the characteristic value of $R$, we turn to the Schmidt relation \cite{2010ApJ...714L.118D}
\begin{equation}
\Sigma_{\rm SFR} = 30\left(\frac{\Sigma_{\rm g}}{\mbox{g\,cm}^{-2}}\right)^{1.4}\,\mbox{M}_{\odot}\,\mbox{yr}^{-1}\,\mbox{kpc}^{-2}
\label{eq:sigmasfr}
\end{equation}
Substituting $\Sigma_{\rm g}= M_{\rm g} / \pi R^2$ into Eq. \ref{eq:sigmasfr} with $M_{\rm g}$ being the galaxy's molecular gas mass, we can express $R$ as
\begin{equation}
R = 2\times 10^{-15}\left(\frac{\Sigma_{\rm SFR}}{\mbox{M}_{\odot}\,\mbox{yr}^{-1}\,\mbox{kpc}^{-2}}\right)^{-1.2}\left(\frac{M_{\rm g}}{\mbox{M}_{\odot}}\right)^{1.7}\,\mbox{kpc}
\label{eq:R}
\end{equation}
The characteristic radius can also be used to determine the characteristic disk height. Following Law et al. \cite{2012ApJ...745...85L}, we adopt $h=0.3R$.

We can further derive the typical value of $M_{\rm g}$ as a function of SFR. Assuming typical dynamical time $\tau_{\rm dyn}\approx10^7$\,yr, the conversion based on Genzel et al. \cite{2010MNRAS.407.2091G} (their Eq. 8) is
\begin{equation}
M_{\rm g} = 5\times 10^9 \left(\frac{\mbox{SFR}}{\mbox{M}_{\odot}\mbox{yr}^{-1}}\right)^{0.73}\,\mbox{M}_{\odot}.
\label{eq:Mg}
\end{equation}
Eqs. \ref{eq:sigmasfr}, \ref{eq:R} and \ref{eq:Mg} together express the typical value of $\Sigma_{\rm g}$ corresponding to a given SFR for starburst galaxies, requiring no other parameter. Note that the above equations represent average relations between parameters. For an example, we take starburst galaxy \emph{Arp\,220}, whose gas density $\Sigma_{\rm g}^{\rm Arp 220}= 5.3$\,g\,cm$^{-2}$ is quoted in \cite{2012ApJ...747L..31S}. The SFR of Arp\,220 is $\sim 100$\,M$_{\odot}$yr$^{-1}$ \cite{2004ApJ...617..966T}, for which our conversion gives $\Sigma_{\rm g} \approx 14$\,g\,cm$^{-2}$, a good estimate given the uncertainties of the measured properties and the variations between starburst galaxies.

\section{Starburst galaxy population}

With the above conversion from SFR to $\Sigma_{\rm g}$ and from $\Sigma_{\rm g}$ to $E_{\rm th}$, we are now able to determine the expected neutrino spectrum for a starburst galaxy with known SFR. We now look at the distribution of starburst galaxy SFR values that allows the derivation of the cumulative neutrino spectrum.

We adopt the starburst galaxy infrared luminosity function by Sargent et al. \cite{2012ApJ...747L..31S}. Sargent et al. find that the luminosity distribution is well described by the Schechter function. We convert their infrared luminosity values at redshift $z\sim1.1$ (see Table 1 of  \cite{2012ApJ...747L..31S}) to SFR based in Kennicutt \cite{1998ARA&A..36..189K}, and fit the derived SFR distribution $\Phi_{\rm SFR}$ with a Schechter function with fixed index $\alpha=1.4$ (see \cite{2012ApJ...747L..31S}), finding parameters $\Phi^{*}=2.2\times10^5$\,Mpc$^{-3}$ and $L^{*}=625$\,M$_{\odot}$yr$^{-1}$.

The SFR distribution also changes with redshift. Following Sargent et al. \cite{2012ApJ...747L..31S}, we take $\Phi^{*}$ to be constant out to $z=1$, and for $z>1$ we adopt scaling with $(1+z)^{-2.40}$. We scale $L^{*}$ with $(1+z)^{2.8}$.

\section{Results}

\subsection{Overall neutrino spectrum at Earth}

The SFR distribution obtained above is converted to the distribution $\Phi_{\rm Eth}$ of cosmic ray $E_{\rm th}$ weighted with galaxy SFR. This latter incorporates the assumption that cosmic ray production is proportional to SFR. Integrating over SFR and redshift gives
\begin{widetext}
\begin{equation}
\Phi_{\rm Eth}(E_{\rm th}') \propto \int_{z}\frac{c\,dz}{H(z)}\int_{\rm SFR} \Phi_{\rm SFR} \, \mbox{SFR} \, E_{\rm th}^{-\Gamma}\, \delta[E_{\rm th}'-E_{\rm th}/(1+z)]\, d\mbox{SFR}
\end{equation}
\end{widetext}
where $\Phi_{\rm SFR}=\Phi_{\rm SFR}(\mbox{SFR},z)$ and $E_{\rm th}=E_{\rm th}(\mbox{SFR})$. $\Phi_{\rm Eth}$ is then converted to neutrino spectrum
\begin{equation}
\frac{dN_{\nu}(E_\nu)}{dE_\nu} = \int_{E_\nu}^{\infty} \Phi_{\rm Eth}(E_{\nu}'/0.05) \left(\frac{E_\nu}{E_{\nu}'/0.05}\right)^{-\Gamma} dE_{\nu}'
\end{equation}
The resulting spectrum is shown in Fig. \ref{figure:spectrum}, for all three magnetic field models considered above. We see that the spectrum deviates from the single power law distribution of cosmic ray acceleration sites ($E^{-\Gamma}$) within the sensitive energy band of IceCube. To describe this softening, we find the empirical fit
\begin{equation}
\frac{dN_{\nu}}{dE_\nu} \propto E_{\nu}^{-\Gamma} \exp\left[-\left(\frac{E_{\nu}}{14 PeV}\right)^{0.43}\right]
\label{eq:fit}
\end{equation}
for our benchmark model (see Fig. \ref{figure:spectrum}). The deviation appears to be slower than exponential. Our results are consistent with that of \cite{2015ApJ...805...95C} who also considered the neutrino spectrum from a population of galaxies, albeit with somewhat different underlying assumptions.

\subsection{Comparison to IceCube spectral fits}

We see that the spectral softening occurs in the IceCube's sensitive energy band for the benchmark magnetic field model. This is in agreement with the prediction of Loeb and Waxman \cite{2006JCAP...05..003L}, who came to this conclusion assuming a population of identical starburst galaxies with characteristic properties (using the same benchmark model for magnetic fields). For the alternative magnetic field models, the spectral softening is above IceCube's band.

If a single power law is fit on the observed distribution in IceCube's sensitive band, the observed $\Gamma_{\rm obs}$ will be greater than expected for a single galaxy below the cutoff energy. Fitting a power law on the neutrino energy band 10\,TeV-10\,PeV for the benchmark model, we find $\Gamma_{\rm fit}\approx2.2$. This spectral fit depends on the energy boundaries considered for the fit.

\subsection{Comparison to GeV gamma-ray diffuse flux by Fermi-LAT}

If neutrinos are generated in $pp$ interactions, there will be a corresponding gamma-ray emission as well. We calculated the expected gamma-ray flux corresponding to our benchmark neutrino flux similarly to the calculations of Murase et al. \cite{2013PhRvD..88l1301M}, taking into account the gamma-ray spectral attenuation due to propagation over cosmic distances. Results are shown in Fig. \ref{figure:spectrum}, in comparison with the diffuse gamma-ray flux detected by Fermi-LAT \cite{2015ApJ...799...86A}. We find that the obtained gamma-ray flux is compatible with the observed diffuse gamma-ray flux, with the gamma-ray flux being $\sim$\,60-100$\%$ of the observed flux around $\sim 100$\,GeV.

\subsection{Constraints on magnetic fields in starburst galaxies}

We see in Fig. \ref{figure:spectrum} that spectral softening occurs within the energy band 10\,TeV-10\,PeV for the benchmark model $B=B_{\rm min}$. For the other two models, the neutrino spectrum essentially follows a power law distribution within 10\,TeV-10\,PeV, and softens only at higher energies. Using the empirical fitting function in Eq. \ref{eq:fit}, we find that the characteristic energy of spectral softening is 14\,PeV for $B=B_{\rm min}$, 110\,PeV for $B=B_{\rm snr}$, and $\gg100$\,PeV for $B=B_{\rm hyd}$. The reconstructed softer spectrum by IceCube therefore supports $B=B_{\rm min}$.

\section{Conclusion}

We characterized the expected high-energy neutrino spectrum from all starburst galaxies by incorporating the galaxies' luminosity function and its cosmic evolution. To accomplish this, we used starburst galaxies' infrared luminosity to derive characteristic properties that determine the neutrino spectrum and luminosity of individual galaxies, such as star formation rate and the magnetic field strength of the interstellar medium. Are conclusions are the following.

\noindent $\bullet$ Spectral decline: Due to the distribution of starburst properties, the expected neutrino spectrum softens within the IceCube energy band ($\sim$\,10\,TeV-10\,PeV) compared to a single power law spectrum, but the decline is not as sharp as the exponential cutoff expected for a single starburst galaxy. We find the empirical fit $dN_\nu/dE \propto E^{-\Gamma}\exp[(E/E_{0})^{\beta}]$ with $\beta\approx 0.43$, representing a slower-than-exponential break.

\noindent $\bullet$ IceCube spectral fit: This has important implications to the spectral fit of IceCube observations. Namely, the observed spectral index within $\sim$\,10\,TeV-10\,PeV range will be softer, $\Gamma_{\rm eff} \sim 2.2$, compared to the injection index $\Gamma=2$ for a single starburst galaxy below the cutoff energy. This observed softening is consistent with the spectral slope obtained by IceCube above $\sim\,$100 TeV, although there may be additional components in the 10-100 TeV range \cite{2015arXiv150900805M}.

\noindent $\bullet$ Fermi-LAT constraints: The expected diffuse gamma-ray background at Earth corresponding to the neutrinos flux is compatible with the spectral constraints from the GeV gamma-ray background observed by Fermi-LAT. The derived gamma-ray flux is $\sim$\,60-100$\%$ of the observed flux.
%
%
Although there is significant tension with the standard blazar interpretation of the diffuse gamma-ray background, the fact that the two diffuse fluxes are compatible may further corroborate starburst galaxies as the potential origin of the observed cosmic high-energy neutrino flux. Since cosmic rays are likely injected into starburst galaxies by energetic transient sources such as gamma-ray bursts and supernovae, this scenario can be promising for multimessenger transient observations, such as those using gravitational waves or gamma rays \cite{2013CQGra..30l3001B,2014PhRvD..90j1301B}.

\noindent $\bullet$ Magnetic field constraints: We investigated alternative magnetic field models for starburst galaxies with respect to their effect on the expected neutrino spectrum. We find that for the benchmark magnetic field based on the observed synchrotron emission from cosmic ray photons predicts neutrino spectral softening in the energy band of recent IceCube results, while other magnetic field models predict significantly higher neutrino break energies, making the spectrum in the IceCube band essentially an unbroken power law. The confirmation of spectral softening in the $\sim$\,10\,TeV-10\,PeV band compared to at TeV energies by IceCube can therefore provide useful constraints on the interstellar magnetic field strength of starburst galaxies.

We foresee two essential future improvements of our calculations. First, there can be more direct measurements of starburst galaxy properties, such as magnetic fields, for a large number of starburst galaxies, which can provide a more accurate estimate than the derived quantities that we used above. Second, our simple starburst-galaxy model can be improved beyond having a sharp spectral cutoff due to the escape of cosmic rays by a more realistic accounting for the fraction of cosmic rays escaping as a function of energy. These improvements can then be directly incorporated in the rest of the analysis presented above. Finally, the general concept presented here, namely the spread of a spectral feature due to a population of diverse sources, can be similarly applied to source candidates other than starburst galaxies.

\begin{acknowledgments}
The authors thank Chad Finley, Peter M\'esz\'aros, Kohta Murase and Xiang-Yu Wang for their useful feedback, and for Kohta Murase for calculating the expected photon flux. IB and SM are thankful for the generous support of Columbia University in the City of New York and the National Science Foundation under cooperative agreement PHY-1447182.
\end{acknowledgments}

\bibliographystyle{h-physrev}

\begin{thebibliography}{10}

\bibitem{2013Sci...342E...1I}
{IceCube Collaboration},
\newblock Science {\bf 342}, 1 (2013), 1311.5238.

\bibitem{2015arXiv150703991I}
{IceCube Collaboration} {\em et~al.},
\newblock ArXiv e-prints  (2015), 1507.03991.

\bibitem{2015arXiv150704005I}
{IceCube Collaboration} {\em et~al.},
\newblock ArXiv e-prints  (2015), 1507.04005.

\bibitem{2014PhRvL.113j1101A}
M.~G. {Aartsen} {\em et~al.},
\newblock Physical Review Letters {\bf 113}, 101101 (2014), 1405.5303.

\bibitem{2015ApJ...805...95C}
X.-C. {Chang}, R.-Y. {Liu}, and X.-Y. {Wang},
\newblock \apj {\bf 805}, 95 (2015), 1412.8361.

\bibitem{2015ApJ...806...24S}
N.~{Senno}, P.~{M{\'e}sz{\'a}ros}, K.~{Murase}, P.~{Baerwald}, and M.~J.
  {Rees},
\newblock \apj {\bf 806}, 24 (2015), 1501.04934.

\bibitem{2008ApJ...689L.105M}
K.~{Murase}, S.~{Inoue}, and S.~{Nagataki},
\newblock \apjl {\bf 689}, L105 (2008), 0805.0104.

\bibitem{2009ApJ...707..370K}
K.~{Kotera} {\em et~al.},
\newblock \apj {\bf 707}, 370 (2009), 0907.2433.

\bibitem{2003ApJ...586L..33R}
G.~E. {Romero} and D.~F. {Torres},
\newblock \apjl {\bf 586}, L33 (2003), astro-ph/0302149.

\bibitem{2006JCAP...05..003L}
A.~{Loeb} and E.~{Waxman},
\newblock \jcap {\bf 5}, 3 (2006), astro-ph/0601695.

\bibitem{2007ApJ...654..219T}
T.~A. {Thompson}, E.~{Quataert}, and E.~{Waxman},
\newblock \apj {\bf 654}, 219 (2007), astro-ph/0606665.

\bibitem{2011ApJ...734..107L}
B.~C. {Lacki}, T.~A. {Thompson}, E.~{Quataert}, A.~{Loeb}, and E.~{Waxman},
\newblock \apj {\bf 734}, 107 (2011), 1003.3257.

\bibitem{2014JCAP...09..043T}
I.~{Tamborra}, S.~{Ando}, and K.~{Murase},
\newblock \jcap {\bf 9}, 43 (2014), 1404.1189.

\bibitem{2012ApJ...744...95R}
B.~E. {Robertson} and R.~S. {Ellis},
\newblock \apj {\bf 744}, 95 (2012), 1109.0990.

\bibitem{1999A&A...351..459C}
E.~{Cappellaro}, R.~{Evans}, and M.~{Turatto},
\newblock \aap {\bf 351}, 459 (1999), astro-ph/9904225.

\bibitem{1995PhRvL..75..386W}
E.~{Waxman},
\newblock Physical Review Letters {\bf 75}, 386 (1995), astro-ph/9505082.

\bibitem{2002Natur.416..823E}
R.~{Enomoto} {\em et~al.},
\newblock \nat {\bf 416}, 823 (2002).

\bibitem{2010PhRvL.104j1101A}
A.~A. {Abdo} {\em et~al.},
\newblock Physical Review Letters {\bf 104}, 101101 (2010), 1002.3603.

\bibitem{2013PhRvD..88l1301M}
K.~{Murase}, M.~{Ahlers}, and B.~C. {Lacki},
\newblock \prd {\bf 88}, 121301 (2013), 1306.3417.

\bibitem{2015arXiv150900805M}
K.~{Murase}, D.~{Guetta}, and M.~{Ahlers},
\newblock ArXiv e-prints  (2015), 1509.00805.

\bibitem{2006ApJ...645..186T}
T.~A. {Thompson}, E.~{Quataert}, E.~{Waxman}, N.~{Murray}, and C.~L. {Martin},
\newblock \apj {\bf 645}, 186 (2006), astro-ph/0601626.

\bibitem{2014ApJ...780..182M}
J.~{McBride}, E.~{Quataert}, C.~{Heiles}, and A.~{Bauermeister},
\newblock \apj {\bf 780}, 182 (2014), 1310.1957.

\bibitem{2012ApJ...747L..31S}
M.~T. {Sargent}, M.~{B{\'e}thermin}, E.~{Daddi}, and D.~{Elbaz},
\newblock \apjl {\bf 747}, L31 (2012), 1202.0290.

\bibitem{1998ARA&A..36..189K}
R.~C. {Kennicutt}, Jr.,
\newblock \araa {\bf 36}, 189 (1998), astro-ph/9807187.

\bibitem{2010MNRAS.407.2091G}
R.~{Genzel} {\em et~al.},
\newblock \mnras {\bf 407}, 2091 (2010), 1003.5180.

\bibitem{2010ApJ...714L.118D}
E.~{Daddi} {\em et~al.},
\newblock \apjl {\bf 714}, L118 (2010), 1003.3889.

\bibitem{2012ApJ...745...85L}
D.~R. {Law} {\em et~al.},
\newblock \apj {\bf 745}, 85 (2012), 1107.3137.

\bibitem{2004ApJ...617..966T}
D.~F. {Torres},
\newblock \apj {\bf 617}, 966 (2004), astro-ph/0407240.

\bibitem{2015ApJ...799...86A}
M.~{Ackermann} {\em et~al.},
\newblock \apj {\bf 799}, 86 (2015), 1410.3696.

\bibitem{2013CQGra..30l3001B}
I.~{Bartos}, P.~{Brady}, and S.~{M{\'a}rka},
\newblock Classical and Quantum Gravity {\bf 30}, 123001 (2013), 1212.2289.

\bibitem{2014PhRvD..90j1301B}
I.~{Bartos} and S.~{M{\'a}rka},
\newblock \prd {\bf 90}, 101301 (2014), 1409.1217.

\end{thebibliography}

\end{document}